\documentclass[preprint,amsmath,amssymb,superscriptaddress]{revtex4-1}
\usepackage{graphicx}% Include figure files
\usepackage{color}
\usepackage{amsfonts}

\begin{document}
\title{Unveiling and veiling  a Schr{\"o}dinger cat state from the vacuum}
%\title{The strange case of the Schr{\"o}dinger and Chessur cats:\\ converting quantum cat states from virtual to real and vice versa}
\author{Roberto Stassi} \email{rstassi@unime.it}
\affiliation{Dipartimento di Scienze Matematiche e Informatiche,Scienze Fisiche e Scienze della Terra, Universit\`a di Messina, I-98166 Messina, Italy}
\affiliation{Theoretical Quantum Physics Laboratory,\\ RIKEN Cluster for Pioneering Research, Wako-shi, Saitama 351-0198, Japan}
\author{Mauro Cirio}
\affiliation{Graduate School of China Academy of Engineering Physics,\\ Haidian District, Beijing, 100193, China}
\author{Ken Funo}
\affiliation{Theoretical Quantum Physics Laboratory,\\ RIKEN Cluster for Pioneering Research, Wako-shi, Saitama 351-0198, Japan}
\author{Neill Lambert}
\affiliation{Theoretical Quantum Physics Laboratory,\\ RIKEN Cluster for Pioneering Research, Wako-shi, Saitama 351-0198, Japan}
\author{Jorge Puebla}
\affiliation{CEMS, RIKEN, Saitama 351-0198, Japan}
\author{Franco Nori}
\affiliation{Theoretical Quantum Physics Laboratory,\\ RIKEN Cluster for Pioneering Research, Wako-shi, Saitama 351-0198, Japan}
\affiliation{RIKEN Center for Quantum Computing, Wako-shi, Saitama 351-0198, Japan}
\affiliation{Physics Department, The University of Michigan,\\ Ann Arbor, Michigan 48109-1040, USA.}

\date{\today}

\begin{abstract}
Deep in the ultrastrong light-matter coupling regime, it has been predicted that the ground state of a two-level atom interacting with a cavity mode takes the form of a ``virtual'' Schr{\"o}dinger cat entangled state  between light and matter. We propose a method to convert this Schr{\"o}dinger cat state from virtual to real, and back again, by driving the atom with optimally chosen pulses. Our system consists of a four-level atom,  with two of these levels ultrastrongly coupled to a cavity mode. We show that the Schr{\"o}dinger cat state can be converted between virtual and real by making use of either an ideal ultrafast pulse or a multi-tone $\pi$-pulse. In addition to allowing us to observe these unusual virtual states this method could also be used to generate entangled cat states on demand for quantum information processing.
\end{abstract}
\maketitle
\section*{Introduction}
In the famous novel by Lewis Carroll, Alice's Adventures in Wonderland, the enigmatic Cheshire Cat appears to Alice sometimes only showing its smile, and other times showing the whole head or body. In this article we describe a phenomenon that resembles the one narrated by Lewis Carroll: a Schr{\"o}dinger cat \cite{scully1999,agarwal2012,ralph2003,gilchrist2004, liu2004generation, wineland2013, mirrahimi2016, chen2021, ku2020, Zhou2021} that behaves like the Cheshire Cat in that it appears and disappears, leaving only part of itself visible to outside observers. To distinguish our phenomenon from the well-known ``quantum Cheshire effect'' \cite{aharonov2013, denkmayr2014, duprey2018,aharonov2021}, we name it after the more modern Chessur cat \cite{TimBurton}.

One of the most fascinating features of quantum mechanics is that the vacuum is filled with virtual particles. In extreme physical conditions these particles can be promoted to real at the expense of a high energy field. For example, Hawking radiation is predicted to be emitted from outside the event horizon of a black hole \cite{hawking1974, Nation:2012ka, scully2018, wang2019}. This phenomenon can be intuitively understood by considering that the Hamiltonian describing a field inside and outside the horizon has an interaction contribution  containing only counter-rotating terms. These terms generate an entangled vacuum between the field inside and outside the horizon. Then, the virtual particles outside the horizon can be promoted to real at the expense of the gravitational field.

Using superconducting circuits \cite{PhysRevLett.55.1543, PhysRevLett.55.1908,  nakamura1999, blais2004cavity, you2011, krantz2019quantum, martinis2020, kjaergaard2020superc, stassi2020}, it is possible to generate extreme physical conditions that allow us to play with the vacuum in a laboratory setting. In particular, it was shown that the interaction energy of a superconducting flux qubit coupled to a cavity field can be larger than the cavity or qubit bare energies \cite{Niemczyk:2010gv, Yoshihara:2017bia, kockum2019g, forn2019}. In this ultrastrong coupling regime the counter-rotating terms \cite{braak2019} in the interaction Hamiltonian modify the vacuum, generating virtual excitations. In contrast to Hawking radiation, the rotating-terms in the interaction Hamiltonian are still present and together with the counter-rotating terms give rise, deep in the ultrastrong coupling regime, to a Schr{\"o}dinger cat light-matter entangled state \cite{Ashhab:2010eh, nataf2010, stassi2018long}. 

\begin{figure}[hbt]
  \includegraphics[scale=0.25]{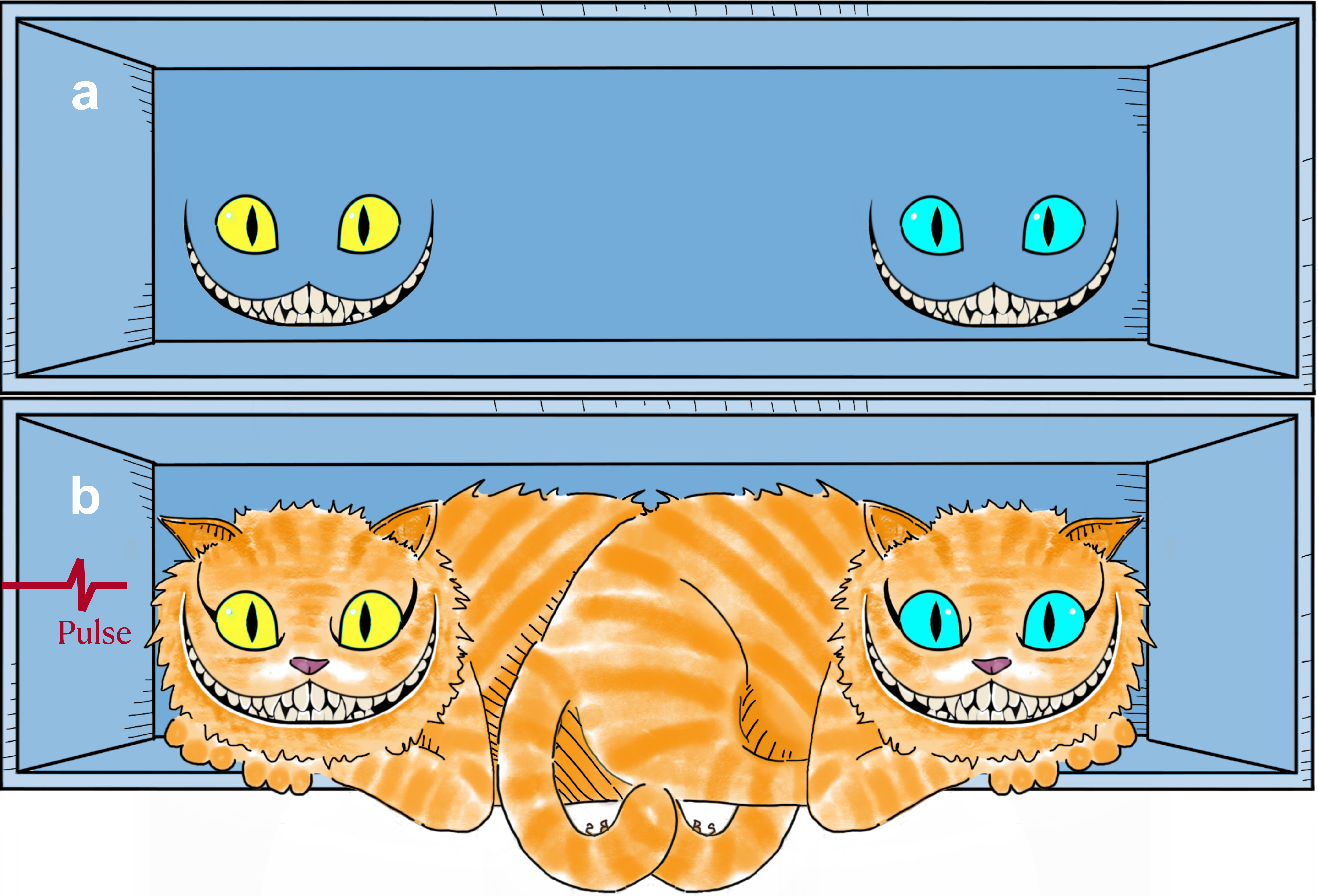}
  \caption{The ground state of a two-level system and a cavity deep in the ultrastrong light-matter coupling regime has been predicted to be a Schr{\"o}dinger cat light-matter entangled state.  \textbf{a} In this system, an observer, opening the box, cannot see the cat because it is a virtual bound state.  However, its presence can be observed: e.g., by measuring the pressure generated by the cat state on the walls of the cavity (in analogy with the Chessur cat, this indirect measurement corresponds to seeing  its smile). \textbf{b} Sending a pulse with a properly chosen frequency to the atom, the cat becomes visible and, opening the box, the external observer will find the cat displaced either to the left or to the right. A second pulse, equal to the first one, will again hide the cat in the ground state.}\label{Fig1}
\end{figure}

In Figure \ref{Fig1}, we show a cartoon representation of the cat state as a superposition of left and right displaced Chessur cats, as an analogy to the  photonic contribution of the ground state when the light-matter interaction is in the ultrastrong coupling regime \cite{Irish:2005ko}. Increasing the light-matter coupling $\lambda$, increases the displacement of these cat states \cite{Ashhab:2010eh}.
This cat state is not ``visible'' (measurable), because it is made of virtual excitations. For this reason, we say that the cat state is ``virtual''.  In analogy with the Chessur Cat smile, see the illustration in Fig.\,\ref{Fig1}a, even if the cat state is invisible, we can still probe it \cite{garziano2014, lolli2015, cirio2017}, because virtual photons generate a pressure to the external environment, so while the whole cat is not visible, we can still infer its presence. 

In this article we theoretically show that the entire Schr{\"o}dinger cat state can be converted from ``virtual'' to ``real'' by swapping the atomic states from ones that interact with the cavity to other ones that do not.  The swap is implemented by driving the atom with an ideal ultrafast mono-tone pulse or with a multi-tone $\pi$-pulse \cite{Stassi:2013jw, di2017feynman, chen2020}. Note that we never drive the cavity. Again in analogy with the Chessur Cat (Fig.\,\ref{Fig1}b), after the pulse, the cat is ``visible'' and an external observer can find it displaced either to the left or to the right. Using an identical pulse, the process can be inverted and the cat becomes again ``invisible''.

\section*{Results}

\textbf{Model}. Our system consists of a 4-level atom, $\{\vert g'\rangle, \vert e'\rangle,\vert g\rangle,\vert e\rangle\}$, and a cavity mode with frequency $\omega_{\rm c}$ \cite{stassi2012delayed}. The transition $\vert g\rangle\leftrightarrow\vert e\rangle$ is ultrastrongly coupled to the cavity. 
The Hamiltonian of the total system is $\hat{H}= \hat{H}_{\rm R}+\hat{H'}+\hat D$, where the Rabi Hamiltonian 
\begin{equation}
  \hat{H}_{\rm R}= \omega_{\rm q}\vert e\rangle\langle e\vert+\omega_{\rm c}\,\hat a^\dagger \hat a+\lambda\,\hat\sigma_x(\hat a+\hat a^{\dagger})\,,
\end{equation}
and $\hat{H'}=\omega_{\rm q}\,\vert e'\rangle\langle e'\vert$ describe respectively the dynamics of the interacting and non-interacting states $(\hbar=1)$. The energy difference between the latter states is defined by $\hat D=\omega_{\rm g}(\vert e\rangle\langle e\vert+\vert g\rangle\langle g\vert)$, $\hat\sigma_{\rm x}=\vert e\rangle\langle g\vert+\vert g\rangle\langle e\vert$ is a Pauli operator, and $\hat a$ is the annihilation operator for the cavity mode. We set $\omega_{\rm c}=\,\omega_{\rm q}/2$ and $\lambda>0.5\,\omega_{\rm q}$.

\begin{figure}[hbt]
  \includegraphics[scale=0.4]{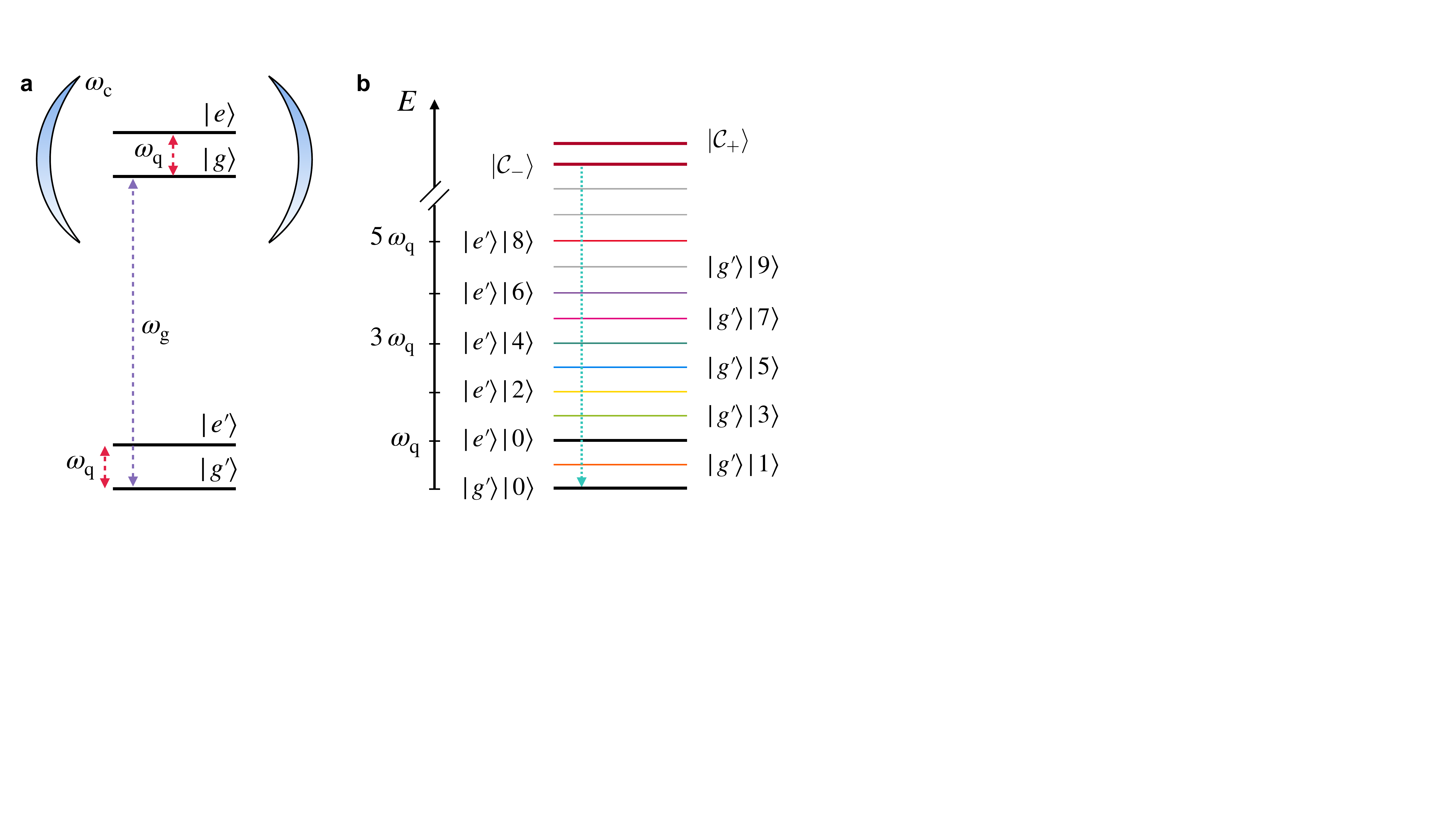}
  \caption{\textbf{a} Four-level atomic system: states $\vert g\rangle$ and $\vert e\rangle$ interact ultrastrongly with a cavity mode. Here $\omega_{\rm g}>\omega_{\rm q}$, implying that interacting levels are higher in energy with respect to non-interacting levels. \textbf{b} Eigenstates of the full Hamiltonian $\hat H$.  The parameters are chosen so that  the atomic states $\vert g'\rangle$ and even number of photons are degenerate with the atomic states $\vert e'\rangle$ and odd number of photons.  $\vert\mathcal{C}_- \rangle$ and $\vert\mathcal{C}_+ \rangle$ are the ground state and first excited state of the Rabi Hamiltonian, and the vertical dashed arrow indicates the ultrafast pulse frequency $\omega_{\rm p}$.}\label{fig2}
\end{figure}

Figure \ref{fig2}a shows the bare atomic states, distinguishing those which do not interact $\{\vert g'\rangle, \vert e'\rangle\}$ and those that do interact $\{\vert g\rangle,\vert e\rangle\}$ with the cavity mode.  We note that for $\omega_{\rm g}>\omega_{\rm q}$ the state $\vert g'\rangle\vert 0\rangle$ is the true ground state of the system, where $\vert 0\rangle$ is the zero photon state, while the ground state of the Rabi Hamiltonian corresponds to an excited state. However, the validity and the significance of the results do not change with this fact.

Figure \ref{fig2}b shows the eigenstates of the total Hamiltonian $\hat H$. For $\lambda\gg0.5\,\omega_{\rm c}$ the atomic states $\{\vert g\rangle,\vert e\rangle\}$ hybridize with the photonic states, giving rise to the  entangled Schr{\"o}dinger cat states, $\vert\mathcal{C}_- \rangle \approx1/\sqrt{2}(\vert+\rangle\vert -\alpha\rangle-\vert -\rangle\vert + \alpha\rangle)$ and $\vert\mathcal{C}_+\rangle\approx1/\sqrt{2}(\vert+\rangle\vert-\alpha\rangle+\vert -\rangle\vert+\alpha\rangle)$. Here, $\vert\pm \rangle=1/\sqrt{2}(\vert e \rangle\pm\vert g \rangle)$ are the eigenstates of $\hat\sigma_{\rm x}$, and $\vert\pm \alpha\rangle=D(\pm\alpha)\vert 0\rangle$ are photonic coherent states of light with positive and negative displacement $\alpha$ \cite{Irish:2005ko}. 
%Because of the dressing, the states $\vert\mathcal{C}_- \rangle$ and $\vert\mathcal{C}_+ \rangle$ are lower in energy with respect to the bare atomic states $\vert g\rangle$ and $\vert e\rangle$. 
The coloured lines in Fig.\,\ref{fig2}b are Fock states associated with the non-interacting sub-space for when the atomic system is in the states $\vert g'\rangle$ and $\vert e'\rangle$.
 
\begin{figure}[hbt]
  \includegraphics[scale=0.4]{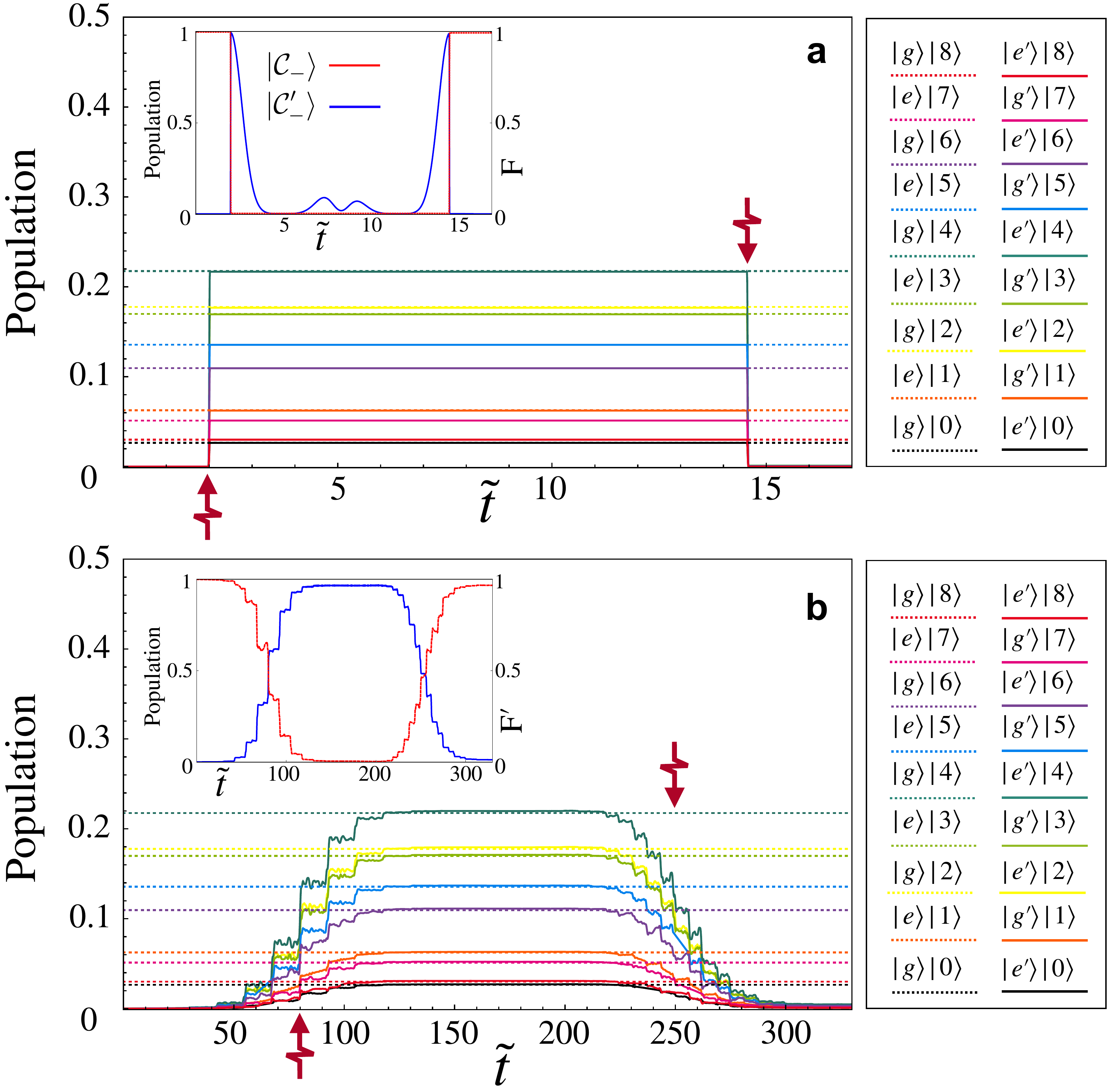} 
  \caption{Dynamics  of the Fock states populations associated with the atomic states $\vert g'\rangle$ and $\vert e'\rangle$ (solid lines) when  the atom  is subject to an an ultrafast pulse, \textbf{a}, or a multi-tone pulse, \textbf{b}. These can be compared with the Fock states associated with $\vert g\rangle$ and $\vert e\rangle$ (dotted lines) when the state of the system is $\vert\mathcal{C}_-\rangle$.  The inset in \textbf{a} shows the population for the $\vert\mathcal{C}_-\rangle$ state (red curve) and the fidelity F for the $\vert\mathcal{C}_-'\rangle$ state (blue curve).  Inset in \textbf{b} represents the population for the $\vert\mathcal{C}_-\rangle$ state (red curve) and the fidelity F' (eliminating relative phases) for the $\vert\mathcal{C}_-'\rangle$ state (blue curve). Parameters in \textbf{a} are $\omega_{\rm g}=9\,\omega_{\rm q}$, $\lambda=\omega_{\rm q}$; in \textbf{b} are $\omega_{\rm g}=6.7\,\omega_{\rm q}$, $\lambda=\omega_{\rm q}$. }\label{Fig3}
\end{figure}

\textbf{Entanglement swapping and conversion}. We now present a swapping procedure in which the entanglement in the dressed light-matter ground state is mapped into entanglement in the non-interacting sector, i.e. for large coupling $\lambda/\omega_{\rm q}>1$, 
 \begin{eqnarray}
  \vert\mathcal{C}_-\rangle \approx\frac{1}{\sqrt{2}}(&\vert+\rangle &\vert-\alpha\rangle-\vert -\rangle\vert+\alpha\rangle)\\
  &\Downarrow &\qquad\qquad\Downarrow\nonumber \\
  \vert\mathcal{C}_-'\rangle \approx\frac{1}{\sqrt{2}}(&\vert+'\rangle &\vert-\alpha\rangle-\vert -'\rangle\vert+\alpha\rangle)\,,
\end{eqnarray}
where, $\vert\pm '\rangle=1/\sqrt{2}(\vert g' \rangle\pm\vert e '\rangle)$.  The swap is realized by inducing atomic transitions $\{\vert g\rangle,\vert e\rangle\}\to\{\vert e'\rangle,\vert g'\rangle\}$, using an ultrafast ideal pulse sent to the atom (Fig.\,\ref{Fig3}a). 
 We also prove that a multi-tone $\pi$-pulse, which may be more physically realistic for some realizations, can efficiently lead to the same result (Fig.\,\ref{Fig3}b).
 
  Figure \ref{Fig3} displays the dynamical evolution of the Fock state populations for the states $\vert g',n\rangle$ and $\vert e',n\rangle$ (solid lines), which can be compared with the Fock states associated with $\vert g,n\rangle$ and $\vert e,n\rangle$ (dotted lines)  representing the bare populations of the state $\vert\mathcal{C}_-\rangle$. 
  
  The general procedure is as follows: first, we prepare the system in the dressed state $\vert\mathcal{C}_-\rangle$. Now, the cat is in the box, but it is not visible because it is generated as a ``virtual state'' by the light-matter interaction, see illustration in Fig.\,\ref{Fig1}a. In Figure \ref{Fig3}a, at $\tilde{t}_0=2$ (up red arrow), with $\tilde{t}=\omega_{\rm q} t/2\pi$, an ultrafast pulse depopulates state $\vert\mathcal{C}_-\rangle$ and populates Fock states associated with $\vert g'\rangle$ and $\vert e'\rangle$ (solid lines). The latter match exactly the state associated with $\vert g\rangle$ and $\vert e\rangle$ (dotted lines) before the pulse was sent. This proves that the cat state now is an entangled state with the sub-space $\{\vert g'\rangle,\vert e'\rangle\}$. The photonic contribution before and after the pulse remains almost unchanged, as shown in Eqs.(1, 2). The cat is now visible to an external observer, as illustrated in Fig.\,\ref{Fig1}b, because the photons are real and can be detected. After the first pulse is sent at $\tilde t_0$, superpositions within the state experience a phase shift \cite{ridolfo2011all, di2011interference}, see inset Fig.\,\ref{Fig3}a. At $\tilde{t}=\tilde{t}_0+\tilde{t}_{\rm d}$ (down red arrow), with $\tilde{t}_{\rm d}=\omega_{\rm q}/\omega_{\rm c}$,  the state is again in phase with the corresponding light-matter ground state $\vert\mathcal{C}_-\rangle$ and a second pulse equal to the first brings back the cat into its original ``virtual form'' encoded in the ground state of the Rabi Hamiltonian  $\vert\mathcal{C}_-\rangle$. Again, for an external observer, the box is now empty, even if the cat is inside ``lying down'' in the ground-state of the system. The ultrafast pulse is given by 
\begin{equation}
  \hat{H}_{\rm p}=\varepsilon_{\rm p}\cos (\omega_{\rm p}( t-t_0))(\hat\sigma_{\rm eg'}+\hat\sigma_{\rm ge'}+H.c.)\,,
\end{equation}
with $\hat\sigma_{\rm mn}=\vert m\rangle\langle n\vert$, $\varepsilon_{\rm p}=\pi\varepsilon\,\exp(-( t- t_0)^2/2A^2_{\rm p})/A_{\rm p}\sqrt{2\pi}$, $\varepsilon=1/2$, $\omega_{\rm p}=\omega_{\mathcal{C}_-}-\omega_{\rm g'}$ and $A_{\rm p}=0.01/\omega_{\rm q}$. In this simulation no dissipation is taken into account. 
 
Figure \ref{Fig3}b shows the dynamical evolution of the system when a multi-tone $\pi$-pulse is sent to the atom and the state of the system is initially $\vert \mathcal{C}_-\rangle$. Here, we consider more experimentally realistic conditions, taking into account the dipole moment of the transition $\vert g\rangle\leftrightarrow\vert e\rangle$ in the Hamiltonian of the pulse, and including environmental effects. The pulse is described by 
\begin{equation}
  \hat{H}_{\rm p}=\varepsilon_{\rm p}\sum_i^N\cos (\omega_{\rm p_{\rm i}}(t-t_0))(\hat\sigma_{\rm eg'}+\hat\sigma_{\rm ge'}+\hat\sigma_{\rm eg}+H.c.)\,,
\end{equation}
where $\omega_{\rm p_{\rm i}}=\omega_{\mathcal{C}_-}-\omega_{\rm i}$, $\omega_{\rm i}$ are the frequencies of $\vert g', \,n\rangle$ and $\vert e',2\,n\rangle$, and $\varepsilon=1$. Notice that the transitions considered in the Hamiltonian describing the pulses satisfy the optical selection rules for a flux qubit in its optimal point \cite{Liu:2005eo}. 

In our numerical simulations we modelled the dissipation using the  master equation \cite{stassi2018long}, $ \dot{\hat\rho}=-i\left[\hat {H}_{\rm S},\hat\rho\right]+\sum_k\sum_{m,\,n>m}\Gamma^{(k)}_{mn}\mathcal{D}\left[\vert m\rangle\langle n\vert\right]\hat\rho,$ where $\hat{H}_{\rm S}=\hat{H}+\hat{H}_{\rm p}$, $\mathcal{D}[\hat O]\hat\rho=(2\hat O\hat\rho\,\hat O^\dagger-\hat O^\dagger\hat O\hat\rho-\hat\rho\,\hat O^\dagger\hat O)/2$ is the Lindblad superoperator, $\vert m\rangle$ are eigenstates of $\hat{H}$, $\hat\rho$ is the density matrix of the system, $\Gamma^{(k)}_{mn}=\gamma^{(k)}\vert \langle m\vert\hat{S}^{(k)}\vert n\rangle\vert^2$, $\hat{S}^{(k)}\in\{\hat\sigma_{\rm eg'},\hat\sigma_{\rm ge'},\hat\sigma_{\rm eg}, \hat a+\hat a^\dagger\}$, and $\gamma^{(k)}$ are the relaxation rates with $\gamma^{(1,2,3)}=10^{-4}\,\omega_{\rm_q}$ and $\gamma^{(4)}=10^{-5}\,\omega_{\rm_q}$. At $\tilde t=80$ (up red arrow), the first pulse is sent to the atom, gradually populating the non-interacting sector until the full entanglement is swapped. At $\tilde t=250$ (down red arrow), a second pulse equal to the first restores (but not perfectly because of dissipation) the original $\vert\mathcal{C}_-\rangle$ state.  We note that, if the parameters are chosen to allow the interacting sector to have lower energy with respect to the non-interacting one, dissipation would automatically restore the initial state within the relaxation time associated with the atom. The insets in Fig.\,\ref{Fig3} show the population of the $\vert\mathcal{C}_-\rangle$ state and the fidelity between the state of the system and the state given by $\vert\mathcal{C}_{\rm ideal}\rangle=(\hat\sigma_{\rm e'g}+\hat\sigma_{\rm g'e})\vert\mathcal{C}_-\rangle$. The fidelity in the inset of Fig.\,\ref{Fig3}b is calculated using the absolute value of the density matrices to eliminate the relative phase.

\begin{figure}[hbt]
  \includegraphics[scale=0.4]{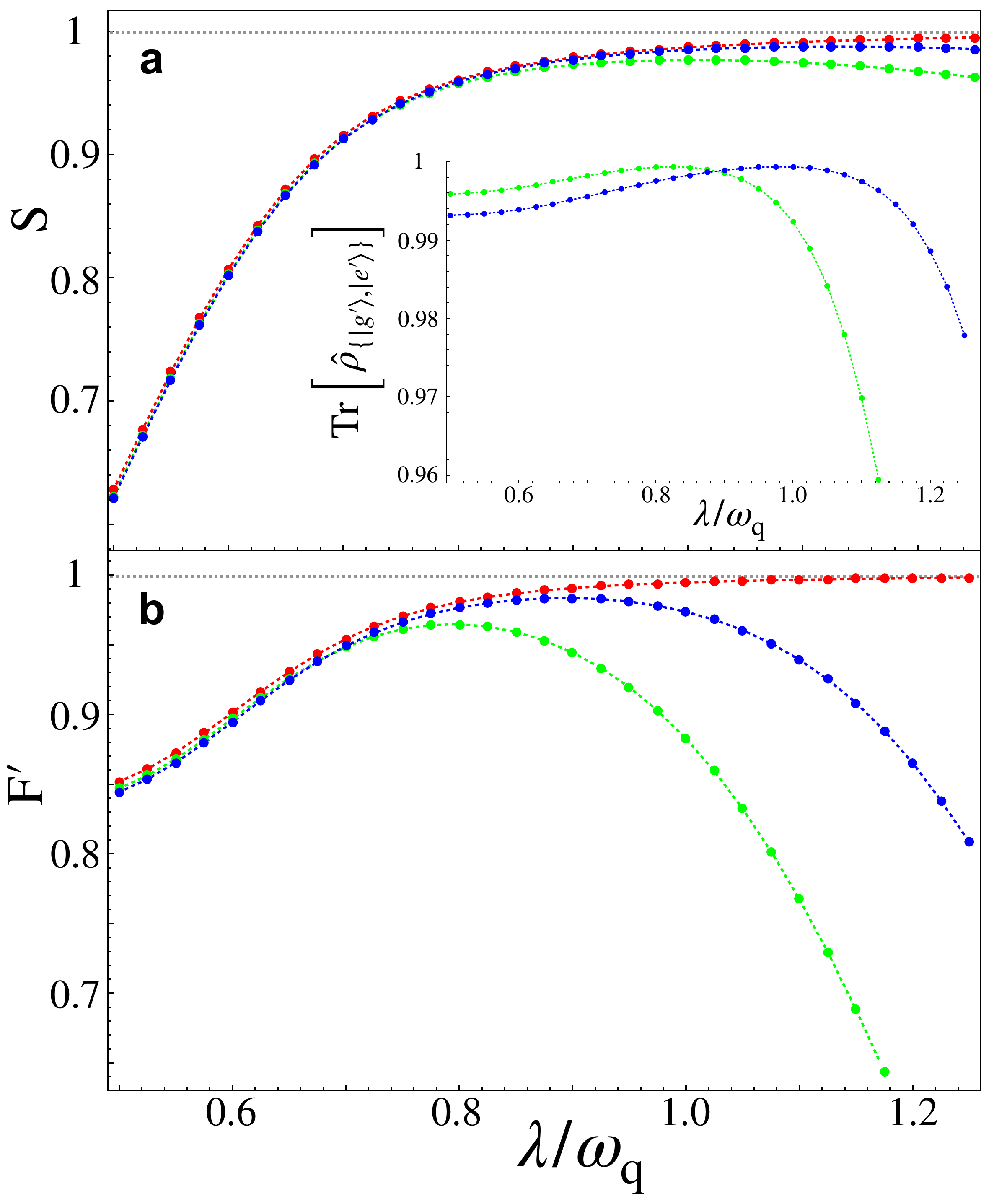}
  \caption{{\bf a} Entanglement entropy S between the field and the atomic subspace $\{\vert g'\rangle,\vert e'\rangle\}$ after a multi-tone $\pi$-pulse with seven frequencies (green dots) and nine frequencies (blue dots) is sent to the atom.  It is possible to compare this to the entropy relative to the subspace $\{\vert g\rangle,\vert e\rangle\}$, when the state of the system is $\vert\mathcal{C}_-\rangle$ (red dots). The inset shows the trace of the density matrix $\hat\rho_{\{\vert g'\rangle, \vert e'\rangle\}}$ for the subspace $\{\vert g'\rangle,\vert e'\rangle\}$, after a multi-tone $\pi$-pulse with seven (green dots) and nine (blue dots) frequencies is sent. Here $\omega_{\rm g}=9\,\omega_{\rm q}$. {\bf b} Fidelity F' (eliminating relative phases) between an ideal cat state entangled with the atom and the system state after a $\pi$-pulse with seven (green dots) and nine frequencies (blue dots) is sent to the atom. The ideal cat state is generated choosing $\alpha=\lambda/\omega_{\rm c}$.}\label{Fig4}
\end{figure}

 \textbf{Analysis}. To quantify the performance of this protocol, we will now calculate the amount of entanglement  that can be transferred from the initial state $\vert \mathcal{C}_-\rangle$ to the state $\vert\mathcal{C}_-'\rangle$. In Figure \ref{Fig4}a, we show the von Neumann entropy \cite{horodecki2009} calculated after sending a multi-tone $\pi$-pulse with seven (green dots) and nine (blue dots) frequencies, and after that we post-select onto the $\{\vert g'\rangle,\vert e'\rangle\}$ subspace.  This approach is justified by the fact that, after the swapping procedure, the residual populations in the $\vert g\rangle$ and $\vert e\rangle$ states are very low (see inset in Fig.\,\ref{Fig4}a). These results can be compared with the entropy calculated for the subsystem $\{\vert g\rangle,\vert e\rangle\}$, when the state of the system is $\vert\mathcal{C}_-\rangle$ (red dots). Note that for couplings in the range $0.5<\lambda/\omega_{\rm q}<0.8$, going from seven to nine pulses does not make any difference. However, for $\lambda/\omega_{\rm q}>0.8$, the number of photons in the ground state increases, and using nine frequencies in the pulse outperforms using seven frequencies. Choosing a target entangled cat state with $\alpha=\lambda/\omega_{\rm c}$ \cite{stassi2018long}, Fig.\,\ref{Fig4}b displays the fidelity for the state of the system after sending the multi-tone $\pi$-pulse with seven (green dots) and nine (blue dots) frequencies. These can be compared with the fidelity for the state $\vert \mathcal{C}_-\rangle$ (red dots).  As before, for large coupling, increasing the number of frequencies in the pulse increases the fidelity.
%$E=-\sum_i\lambda_i\log\lambda_i$ where $\{\lambda_i\}$ are eigenstates of the system density matrix.
 
\section*{Discussion}

In this article, we proposed a protocol to unveil the virtual entangled Schr{\"o}dinger cat state that it is predicted to be the ground state of a two-level system ultrastrongly coupled to a cavity mode.  To unveil and make the cat visible and measurable the entangled partner is swapped using an ideal ultrafast pulse or a multi-tone $\pi$-pulse. With this proposal, not only it is possible to explore the ground state of the Rabi Hamiltonian in the ultrastrong coupling regime, but also it can be an efficient way to generate photonic cat states on demand, that can be an important resource for quantum computation and other quantum technologies \cite{joo2011, mirrahimi2014}. Moreover, we believe that this novel mechanism can potentially enable laboratory experiments mimicking virtual particle emission from black holes or accelerating particles (Unruh effect), giving the chance to investigate, in laboratory conditions, the black-hole information paradox \cite{mathur2009}.

It is possible  to realize this proposal using a superconductor flux qubit coupled to a coplanar waveguide or an $LC$ resonator with, i.e., charging energy $E_{\rm C}=4.4\,\rm GHz$, Josephson energy $E_{\rm J}=15\,E_{\rm C}$, $\alpha=0.8$, and an external flux with $f=0.5$ \cite{Mooij:1999ex, devoret2004, clarke2008superc, gu2017}. With these parameters, optical selection rules allow only the transitions $\vert g'\rangle\leftrightarrow\vert e\rangle$, $\vert e'\rangle\leftrightarrow\vert g\rangle$, and $\vert g'\rangle\leftrightarrow\vert e'\rangle$. Other transitions are forbidden \cite{Liu:2005eo}. The transition $\vert g'\rangle\leftrightarrow\vert e'\rangle$ can be engineered to be in the ultrastrong coupling regime, while the higher levels are now the non-interacting subspace, and as a consequence  the cat state we describe in this work would be the true ground state of the total system.
\section*{Acknowledgements}
F.N. is supported in part by:
Nippon Telegraph and Telephone Corporation (NTT) Research,
the Japan Science and Technology Agency (JST) [via
the Quantum Leap Flagship Program (Q-LEAP), the Moonshot R\&D Grant Number JPMJMS2061, and
the Centers of Research Excellence in Science and Technology (CREST) Grant No. JPMJCR1676],
the Japan Society for the Promotion of Science (JSPS)
[via the Grants-in-Aid for Scientific Research (KAKENHI) Grant No. JP20H00134 and the
JSPS–RFBR Grant No. JPJSBP120194828],
the Army Research Office (ARO) (Grant No. W911NF-18-1-0358),
the Asian Office of Aerospace Research and Development (AOARD) (via Grant No. FA2386-20-1-4069), and
the Foundational Questions Institute Fund (FQXi) via Grant No. FQXi-IAF19-06.

%Schr{\"o}dinger cat states are very important for quantum computation, our protocol is also an efficient and fast way to generate this states on demands. 

 %Even if in laboratory it is possible to go beyond the ultrastrong coupling regime \cite{Yoshihara:2017bia}, till now there is no definitive evidence of the presence of Schr{\"o}dinger cat states in the ground..
% With this proposal it could be possible to definitively prove that there are cat states in ground and, consequently,  will be possible to fast generate on demand entangled Schr{\"o}dinger cat states, these are very important for quantum computation and quantum metrology \cite{gilchrist2004Schr{\"o}dinger, mirrahimi2014dynamically, chen2021shortcuts}

\bibliographystyle{naturemag}
\bibliography{SchrodingerCatFreedom}

\newpage

\end{document}